\documentclass[11pt]{article}

\usepackage{times}
\usepackage{fullpage}
\usepackage{graphicx}
\usepackage{amssymb}
\usepackage{amsmath}
\usepackage{colortbl}
\usepackage{color}
\usepackage[bf]{caption}
\usepackage{indentfirst}
\usepackage{url}
\usepackage{caption}
\usepackage{subcaption}
\usepackage{multirow}
\usepackage{threeparttable}
\usepackage{tikz}
\usepackage{bbding}
\usepackage{balance}
\usepackage[all]{xy}
\usepackage{algorithmic}
\usepackage{array}
\usepackage{multirow}
\usepackage{paralist}
\usepackage{xspace}
\usepackage{anyfontsize}
\usepackage[english]{babel}
\usepackage{url}
\definecolor{darkgreen}{RGB}{0,0,180}
\definecolor{darkergreen}{RGB}{0,0,120}
\usepackage{breakurl}
\usepackage[bookmarks=false, colorlinks=true, plainpages=false,  linkcolor=darkgreen,   citecolor=darkgreen, urlcolor=darkergreen, filecolor=darkgreen]{hyperref}

\renewcommand{\vec}[1]{\mathbf{#1}}
\makeatletter
\def\url@leostyle{%
  \@ifundefined{selectfont}{\def\UrlFont{}}%
  {\def\UrlFont{}}%
}
\makeatother	
\DeclareMathOperator*{\argmin}{arg\,min}

\makeatletter
\newcommand{\Spvek}[2][r]{%
  \gdef\@VORNE{1}
  \left(\hskip-\arraycolsep%
    \begin{array}{#1}\vekSp@lten{#2}\end{array}%
  \hskip-\arraycolsep\right)}

\def\vekSp@lten#1{\xvekSp@lten#1;vekL@stLine;}
\def\vekL@stLine{vekL@stLine}
\def\xvekSp@lten#1;{\def\temp{#1}%
  \ifx\temp\vekL@stLine
  \else
    \ifnum\@VORNE=1\gdef\@VORNE{0}
    \else\@arraycr\fi%
    #1%
    \expandafter\xvekSp@lten
  \fi}
\makeatother

\newcommand{\descr}[1]{\vspace{0.25cm} \noindent \textbf{#1}}

\begin{document}
\pagestyle{plain}

\title{\bf What's the Gist? \\Privacy-Preserving Aggregation of User Profiles\thanks{A preliminary version of this paper appears in the Proceedings of ESORICS 2014. This is the full version.}}

\author{Igor Bilogrevic$^1$, Julien Freudiger$^2$, Emiliano De Cristofaro$^3$, Ersin Uzun$^2$\\[1ex]
$^1$ Google, Switzerland$^{\dag}$ $\;\;$ $^2$ PARC, USA $\;\;$ $^3$ UCL, UK$^{\dag}$}
\date{}

\maketitle

\begin{abstract}

Over the past few years, online service providers have started gathering increasing amounts of personal information to build user profiles and monetize them with advertisers and data brokers. Users have little control of what information is processed and are often left with an \emph{all-or-nothing} decision between receiving free services or refusing to be profiled. This paper explores an alternative approach where users only disclose an {\em aggregate model} -- the ``gist'' -- of their data. We aim to preserve data utility and simultaneously provide user privacy. We show that this approach can be efficiently supported by letting users contribute encrypted and differentially-private data to an aggregator. The aggregator combines encrypted contributions and can only extract an aggregate model of the underlying data. We evaluate our framework on a dataset of 100,000 U.S. users obtained from the U.S. Census Bureau and show that (i) it provides accurate aggregates with as little as 100 users, (ii) it generates revenue for both users and data brokers, and (iii) its overhead is appreciably low.

\end{abstract}

\renewcommand{\thefootnote}{}
\footnotetext{$^\dag$Work done while authors were at PARC.}
\renewcommand{\thefootnote}{\arabic{footnote}}

\section{Introduction}\label{sec:introduction}
The digital footprint of Internet users is growing at an unprecedented pace, driven by the pervasiveness of online interactions and large number of posts, likes, check-ins, and content shared everyday. This creates invaluable sources of information that online service providers use to profile users and serve targeted advertisement.
This economic model, however, raises major privacy concerns~\cite{comResPrivacy,privacyConcerns,nyTimesRules} as 
advertisers might excessively track users,
data brokers might illegally market consumer profiles~\cite{spokeoPenalized}, 
and governments might abuse their surveillance power~\cite{prismWashington,prismGuardian} by obtaining datasets collected for other purposes (i.e., monetization).
Consequently, consumer advocacy groups are promoting policies and legislations providing greater control to users and more transparent collection practices~\cite{nyTimesGovern,nyTimesRules}.

Along these lines, several efforts -- such as OpenPDS, personal.com, Sellbox, and Handshake -- advocate a novel, user-centric paradigm: users store their personal information in ``data vaults'', and directly manage with whom to share their data. This approach has several advantages, namely, users maintain data ownership (and may monetize their data), while data brokers and advertisers benefit from more accurate and detailed personal information~\cite{malheiros2013fairly,tunner2013bizarro}. %
Nevertheless, privacy still remains a challenge as users need to trust data vaults operators and relinquish their profiles to advertisers~\cite{carrascal2013your,Hushmail}. 

To address such concerns, the research community has proposed to maintain data vaults on user devices and share data in a privacy-preserving way. 
Prior work can be grouped into three main categories:
(1) serving ads locally, without revealing any information to advertisers/data brokers~\cite{guha2011privad,mohan2013prefetching,toubiana2010adnostic};  
(2) relying on a trusted third party to anonymize user data~\cite{backes2012obliviad,riederer2011sale}; and 
(3) relying on a trusted third party for private user data aggregation~\cite{akkus2012non,chen2013splitx,chen2012towards}. 
Unfortunately, these approaches suffer from several limitations. First, localized methods prevent data brokers and advertisers from obtaining user statistics. Second,
anonymization techniques provide advertisers with significantly reduced data utility and are prone to re-identification attacks~\cite{narayanan2008robust}. Finally, existing private aggregation schemes rely on a trusted third party for differential privacy (e.g., a proxy~\cite{chen2012towards}, a website~\cite{akkus2012non}, or mixes~\cite{chen2013splitx}; also, aggregation occurs after decryption, thus making it possible to link contributions and users.

Motivated by the above challenges, this paper proposes a novel approach to privacy-preserving aggregation of user data. Rather than contributing data \emph{as-is}, users combine their data into \emph{an aggregate model} -- the ``gist.'' Intuitively, users contribute encrypted and differentially-private data to an aggregator that extracts a statistical model of the underlying data (e.g., probability density function of the age of contributing users). Our approach addresses issues with existing work in that it does not depend on a third-party for differential privacy, incurs low computational overhead, and addresses linkability issues between contributions and users. Moreover, we propose a metric to dynamically value user statistics according to their inherent amount of ``valuable'' information (i.e., sensitivity): for instance, aggregators can assess whether age statistics in a group of participants are more sensitive than income statistics. To the best of our knowledge, our solution provides the first privacy-preserving aggregation scheme for personal data monetization.  %

Our contributions can be summarized as follows:
\begin{enumerate}
\item We design a privacy-preserving framework for monetizing user data, where users trade an aggregate of their data instead of actual values.
\item We define a measure of the sensitivity of different data aggregates. In particular, we adopt the information-theoretic Jensen-Shannon divergence~\cite{lin1991divergence} to quantify the distance between the actual distribution of a data attribute, and a distribution that does not reveal actionable information~\cite{feldman1995knowledge}, such as the uniform distribution. 
\item We show how to rank aggregates based on their sensitivity, i.e., we design a dynamic valuation scheme based on how much information an aggregate leaks. 
\end{enumerate}

We evaluate our privacy-preserving framework on a real, anonymized dataset of 100,000 US users (obtained by the Census Bureau) with different types of attributes. Our results show that our framework (i) provides accurate aggregates with as little as 100 participants, (ii) generates revenue for users and data aggregators depending on the number of contributing users and sensitivity of attributes, and (iii) has low computational overhead on user devices (0.3 ms for each user, independently of the number of participants). 
Interestingly, we find that data brokers have an incentive to direct their investments on small groups of users representative of a certain population.
In summary, our approach provides a novel perspective to the privacy-preserving monetization of personal data, and finds a successful balance between data accuracy for advertisers, privacy protection for users, and incentives for data aggregators.

\descr{Paper Organization.} The rest of the paper is organized as follows. Next section introduces the system architecture and the problem statement. Then, Section \ref{sec:monetizing} presents our framework and Section \ref{sec:eval} reports on our experimental evaluation. After reviewing related work in Section \ref{sec:rel-work}, we conclude the paper in Section \ref{sec:concl}.
\section{System Architecture}\label{sec:sys-arch}
This section introduces the problem definition and presents participating entities.

\subsection{Problem Statement}
\begin{figure}[ttt]
\centering
\includegraphics[width=0.8\linewidth]{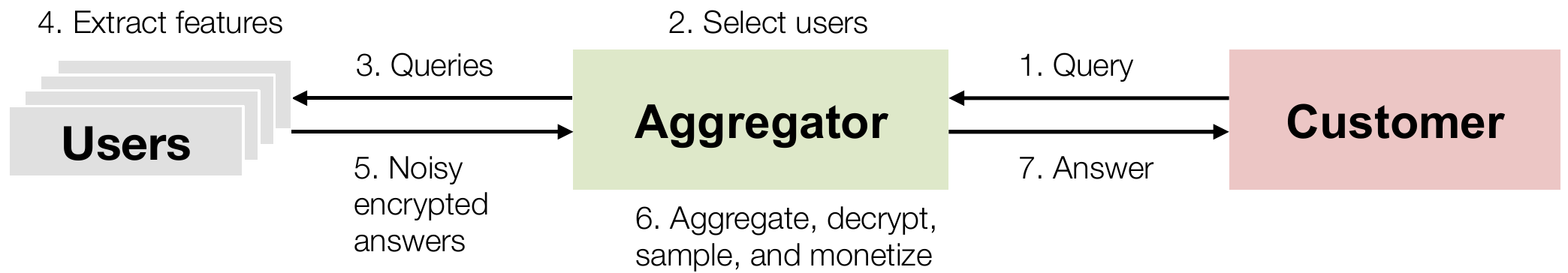}
\vspace{-0.1cm}
\caption{System architecture and basic protocol. Users contribute encrypted profiles to the aggregator. The aggregator combines encrypted profiles and obtains plaintext data models, which it monetizes with customers. }
\label{fig:system-model}
\end{figure}

We consider a system comprised of three entities: A set of users $\mathbb{U} = \{1,\ldots,N\}$, a data aggregator $\mathbb{A}$, and a customer $\mathbb{C}$. The system architecture is illustrated in Fig. \ref{fig:system-model}. 
Customers query the data aggregator for user information, while users contribute their personal information to the data aggregator. The aggregator acts as a proxy between users and customers by aggregating (and monetizing) user data.
The main goal of this paper is to propose practical techniques to aggregate and monetize user personal data in a privacy-preserving way, i.e., without revealing personal information to other users or third parties.

\subsection{System Model}

\descr{Users.} 
We assume that users store a set of personal attributes such as age, gender, and preferences locally. %
Each user $i \in \mathbb{U}$ maintains a profile vector $\vec{p_i} = [x_{i,1},\ldots, x_{i,K}]$, where $x_{i,j} \in \mathcal{D}$ is the value of attribute $j$ and $\mathcal{D}$ is a suitable domain for $j$. For example, if $j$ represents the age of user $i$, then $x_{i,j} \in \{1,\ldots,M_j\}$, $M_j = 120$, and $\mathcal{D} \subset \mathbb{N}$.

In practice, users can generate their personal profiles manually, or leverage profiles maintained by third parties.
Several social networks allow subscribers to download their online profile. A Facebook profile, for example, contains numerous Personally Identifiable Information (PII) items (such as age, gender, relationships, location), preferences (movies, music, books, tv shows, brands), media (photos and videos) and social interaction data (list of friends, wall posts, liked items). %
Following the results of recent studies on user privacy attitudes~\cite{aperjis2012market,carrascal2013your,malheiros2013fairly}, we assume that each user $i$ can specify a privacy-sensitivity value $0\leq \lambda_{i,j} \leq 1$ for each attribute $j$. A large $\lambda_{i,j}$ indicates high privacy sensitivity (i.e., lower willingness to disclose). In practice, $\lambda_{i,j}$ can assume a limited number of discrete values, which could represent the different levels of sensitivity according to Westin's Privacy Indexes \cite{kumaraguru2005privacy}.

We assume that users want to monetize their profiles while preserving their privacy. 
For instance, users may be willing to trade an {\em aggregate} of their online behavior, such as the {\em frequency} at which they visit different {\em categories} of websites, rather than the exact time and URLs. 

Finally, we assume that user devices can perform cryptographic operations consisting of multiplications, exponentiations, and discrete logarithms. %

\descr{Data Aggregator.} A data aggregator $\mathbb{A}$ is an untrusted third-party that performs the following actions: 
(1) it collects encrypted attributes from users, 
(2) it aggregates contributed attributes in a privacy-preserving way, and 
(3) it monetizes users' aggregates according to the amount of ``valuable'' information that each attribute conveys. 

We assume that users and $\mathbb{A}$ sign an agreement upon user registration that authorizes $\mathbb{A}$ 
to access the aggregated results (but not users' actual attributes), 
to monetize them with customers, and
to take a share of the revenue from the sale.
It also binds $\mathbb{A}$ to redistribute the rest of the revenue among contributing users.

\descr{Customer.} We consider a customer $\mathbb{C}$ willing to obtain aggregate information about users and to pay for it. 
$\mathbb{C}$ can have commercial contracts with multiple data brokers. Similarly, a data aggregator can have contracts with multiple customers. %
$\mathbb{C}$ interacts with a data aggregator $\mathbb{A}$ and does not communicate directly with users. $\mathbb{C}$ obtains available attributes, and initiates an aggregation by querying the data aggregator for specific attributes.

\subsection{Applications}
The proposed system model is well-suited to many real-world scenarios, including market research and online tracking use cases. 
For instance, consider a car dealer $\mathbb{C}$ that wants to assess user preferences for car brands, their demographics, and income distributions. A data aggregator $\mathbb{A}$ might collect aggregate information about a representative set of users $\mathbb{U}$ and monetize it with the car dealer $\mathbb{C}$. Companies such as Acxiom currently provide this service, but raise privacy concerns~\cite{nytimesAxciom}. Our solution enables such companies to collect aggregates of personal data instead of actual values and reward users for their participation. 

Another example is that of an online publisher (e.g., a news website) $\mathbb{C}$ that wishes to know more about its online readership~\cite{akkus2012non}. In this case, the aggregator $\mathbb{A}$ is an online advertiser that collects information about online users $\mathbb{U}$ and monetizes it with online publishers. 

Finally, our proposed model can also be appealing to data aggregators in healthcare~\cite{DataCommons}. Healthcare data is often fragmented in silos across different organizations and/or individuals.
An healthcare aggregator $\mathbb{A}$ can compile data from various sources and allow third parties $\mathbb{C}$ to buy access to the data. At the same time, data contributors ($\mathbb{U}$) receive a fraction of the revenue. 
Our approach thwarts privacy concerns and helps with the pricing of contributed data.

\subsection{Threat Model}
In modeling security, we consider both passive and active adversaries. %

\descr{Passive adversaries.} Semi-honest (or honest-but-curious) passive adversaries monitor user communications and try to infer the individual contributions made by other users. For instance, users may wish to obtain attribute values of other users; similarly, data aggregators and customers may try to learn the values of the attributes from aggregated results. A passive adversary executes the protocol correctly and in the correct order, without interfering with inputs or manipulating the final result.

\descr{Active adversaries.} Active (or malicious) adversaries can deviate from the intended execution of the protocol by inserting, modifying or erasing input or output data. For instance, a subset of malicious users may collude with each other in order to obtain information about other (honest) users or to bias the result of the aggregation. To achieve their goal, malicious users may also collude with either the data aggregator or with the customer. Moreover, a malicious data aggregator may collude with a customer in order to obtain private information about the user attributes.

\section{Monetizing User Profiles with Privacy}\label{sec:monetizing}
We outline and formalize the data monetization framework, which consists of a protocol that is executed between users $\mathbb{U}$, a data aggregator $\mathbb{A}$ and a customer $\mathbb{C}$. We first provide an intuitive description and then detail each individual component.

\subsection{High-Level Description}\label{subsec:protocol}
We propose a protocol where users trade personal attributes in a privacy-preserving way, in exchange for (possibly) monetary retributions. 
Intuitively, there are two possible modes of implementations: {\em interactive} and {\em batch}. 

In interactive mode, a customer initiates a query about specific attributes and users. The aggregator selects users matching the query, collects encrypted replies, computes aggregates, and monetizes them according to a pricing function. 

In batch mode, users send their encrypted profile, containing personal attributes, to the data broker. The aggregator combines encrypted profiles, decrypts them, obtains aggregates for each attribute, and ranks attributes based on the amount of ``valuable'' information they provide. A customer is then offered access to specific attributes. %
Without loss of generality, hereafter we describe the interactive mode. 

\descr{Initialization}: The data aggregator $\mathbb{A}$ and users $i \in \mathbb{U}$ engage in a secure key establishment protocol to obtain individual random secret keys $s_j$, where $s_0$ is only known to $\mathbb{A}$ and $s_i$ ($\forall i \in \mathbb{U}$) is only known to user $i$, such that $s_0+s_1+\ldots+s_N = 0$ (this condition is required for the data aggregation described hereafter). Any secure key establishment protocol or trusted dealer can be used in this phase to distribute the secret keys, as long as the condition on their sum is respected. The initialization phase is the same as in~\cite{shi2011privacy}. Each user $i$ generates its profile vector $\vec{p_i} \in \mathcal{D}^K$ containing personal attributes $j \in \{1,\ldots,K\}$.

\begin{enumerate}

\item \textbf{Customer Query}: A customer queries the aggregator. The query contains information about the type of aggregates and users. In practice, it could be formatted as an SQL query. 

\item \textbf{User Selection}: The aggregator selects users based on the customer query. To do so, we consider that users shared some basic information with the aggregator, such as their demographics. Another option is for the aggregator to forward the customer query to users, and let users decide whether to participate or not.  %

\item \textbf{Aggregator Query}: The aggregator forwards the customer's query to the users, together with a public feature extraction function $f$. 

\item \textbf{Feature Extraction}: 
Each user $i$ can optionally execute a public feature extraction function $f: \mathcal{D}^K \rightarrow \mathcal{O}^L$ on $\vec{p_i}$, where $L$ is the dimension of the output feature space  $\mathcal{O}$, thus resulting in a feature vector $\vec{f_i}$.  In our implementation, we consider a simple function that extracts the value of an attribute and its square. %

\item \textbf{Encryption and Obfuscation}: Each user adds noise to $\vec{f_i}$, obtaining $\widehat{\vec{f_i}}$, and encrypts it. Encryption and obfuscation provide strong guarantees both in terms of data confidentiality and differential privacy \cite{dwork2006differential}. Each user sends the encrypted vector $\mathcal{E}(\widehat{\vec{f_i}})$ to $\mathbb{A}$.

\item \textbf{Aggregation, Decryption, and Pricing}: $\mathbb{A}$ combines all $\mathcal{E}(\widehat{\vec{f_i}})$ and decrypts the result, generating a 2-tuple $(V_j,W_j) \in \mathbb{R}^2$ for each attribute $j$. These tuples are used to approximate the probability density function of attributes across users.
$\mathbb{A}$ uses $(V_j,W_j)$ to create a discrete sampled probability distribution function $\text{d}\mathcal{N}_j$ for each attribute $j$. $\mathbb{A}$ then computes a distance measure $d_j=d(\text{d}\mathcal{N}_j,\text{d}\mathcal{U}_j) \in [0,1]$ between $\text{d}\mathcal{N}_j$ and $\text{d}\mathcal{U}_j$, where $\text{d}\mathcal{U}_j$ is a discrete uniform distribution in the interval $[m_j,M_j]$. %
A small/large distance corresponds to an attribute with low/high information ``value'', as described later in the text.

$\mathbb{A}$ determines the cost \emph{Cost}$(j)$ of each attribute $j$ by taking into account both the distances $d_j$, the number of contributing users, and the price per attribute. %

\item \textbf{Answer}: $\mathbb{A}$ sends a set of 2-tuples $\{(d_{\rho_z},Cost(\rho_z))\}_{\rho_z = 1}^K$ to $\mathbb{C}$, which selects  aggregates to purchase. After the purchase, $\mathbb{A}$ obtains a share of the total sale revenue and equally distributes the remainder to users.
\end{enumerate}

\subsection{Detailed Description}\label{subsec:parametric}

We detail the functions and primitives for the aggregation and monetization of user data. 
In this paper, we compute aggregates by estimating the probability density function ($pdf$) of user attributes. We use the Gaussian approximation to estimate $pdf$s for two reasons. 
First, existing work shows that this will lead to precise aggregates with few users. 
The CLT~\cite{marquis1810memoire,rice1995mathematical} states that the arithmetic mean of a sufficiently large number of independent random variables, drawn from distributions of expected value $\mu$ and variance $\sigma^2$, will be approximately normally distributed $\mathcal{N}(\mu, \sigma^2)$. 
Second, a Gaussian $pdf$ $\mathcal{N}$  is fully defined by two parameters and thus we do not need additional coordination among users (after the initialization phase). For information leakage ranking, we use a well-established information-theoretic distance function. %

For conciseness, we focus on the description of privacy-preserving aggregation and pricing (phases 4 to 6, i.e., feature extraction, encryption, aggregation and ranking). %
With respect to the initialization and query forwarding phases (1-3), our method is general enough and can be adapted to any specific implementation.

\subsubsection{Phase 4-5: Feature Extraction and Encryption.}\label{subsec:prof-extr}
Each user $i$ generates a profile vector $\vec{p_i} = [x_{i,1},\ldots, x_{i,K}]$. Each attribute $j$ takes value $x_{i,j} \in \{m_j,\ldots,M_j\}$, where $m_j, M_j \in \mathbb{Z}_p$ are the minimum and maximum value. 
Note that as in~\cite{shi2011privacy}, computations are in cyclic group $\mathbb{Z}_p$ of prime order $p$. The aggregator also chooses a generator $g$ at random, such that $g \in \mathbb{Z}_p$, and $H$ is a Hash function. 
Remember that in practice, a user can derive $\vec{p_i}$ either from an existing online profile (e.g., Facebook) or by manually entering values $x_{i,j}$. In our evaluation, we use values from the U.S. Census Bureau \cite{dataferrett,datagov}.

We consider a simple feature extraction $f$ that consists in providing $x_j$ and computing $x_j^2$. Obviously, other feature extraction method may contribute higher-order moments or simply combine attributes together to obtain richer $x_j$'s.  

To %
guarantee $(\epsilon,\delta)$-Differential Privacy, each user $i$ adds noise $r_{i,j},o_{i,j}$ to attribute values sampled from a symmetric Geometric distribution according to Algorithm 1 in~\cite{shi2011privacy}. In particular, in the following we add noise to both $x_{i,j}$ and $x_{i,j}^2$, as they will be subsequently combined to obliviously compute the parameters of the model that underlies the actual data: %
$$\widehat{x_{i,j}} = x_{i,j} + r_{i,j} \mod{p}$$ and $$\widehat{x_{i,j}}^{(2)} = x_{i,j}^2 + o_{i,j} \mod{p}$$
where $p$ is the prime order~\cite{shi2011privacy}.

With $\widehat{x_{i,j}}$ and $\widehat{x_{i,j}}^{(2)}$, each user generates the following encrypted vectors $(\vec{c_i},\vec{b_i})$:

$\vec{c_i} = \Spvek{c_{i,1};c_{i,2};\vdots;c_{i,K}} = \Spvek{g^{\widehat{x_{i,1}}} H(t)^{s_i};g^{\widehat{x_{i,2}}} H(t)^{s_i};\vdots;g^{\widehat{x_{i,K}}} H(t)^{s_i}}$,
\qquad
$\vec{b_i} = \Spvek{b_{i,1};b_{i,2};\vdots;b_{i,K}} = \Spvek{g^{\widehat{x_{i,1}}^{(2)}} H(t)^{s_i};g^{\widehat{x_{i,2}}^{(2)}} H(t)^{s_i};\vdots;g^{\widehat{x_{i,K}}^{(2)}} H(t)^{s_i}}$

Each user $i$ then sends $(\vec{c_i},\vec{b_i})$ to $\mathbb{A}$. Note that the encryption scheme guarantees that $\mathbb{A}$ is unable to decrypt the vectors $(\vec{c_i},\vec{b_i})$. However, thanks to its own secret share $s_0$, $\mathbb{A}$ can decrypt aggregates as explained hereafter.

\subsubsection{Phase 6: Privacy-Preserving Aggregation and Pricing.}\label{subsec:distr-rank}
To compute the sample mean $\widehat{\mu_j}$ and variance $\widehat{\sigma_j^2}$ without having access to the individual values $\widehat{x_{i,j}},\widehat{x_{i,j}}^{(2)}$ of any user $i$, $\mathbb{A}$ first computes intermediate 2-tuple $(V_j,W_j)$: 
\begin{align*}
V_j &= H(t)^{s_0} \Pi_{i=1}^N c_{i,j} = H(t)^{\sum_{k=0}^N s_k} g^{\sum_{i=1}^N \widehat{x_{i,j}}} = g^{\sum_{i=1}^N \widehat{x_{i,j}}}\\
W_j &= H(t)^{s_0} \Pi_{i=1}^N b_{i,j} = H(t)^{\sum_{k=0}^N s_k} g^{\sum_{i=1}^N \widehat{x_{i,j}}^{(2)}} = g^{\sum_{i=1}^N \widehat{x_{i,j}}^{(2)}}
\end{align*}
To obtain $(\widehat{\mu_j},\widehat{\sigma_j^2})$, $\mathbb{A}$ takes the discrete logarithm base $g$ of $(V_j,W_j)$:
\begin{align*}
\widehat{\mu_j} &= \frac{\log_g(V_j)}{N} = \frac{\sum_{i=1}^N \widehat{x_{i,j}}}{N}\\
\widehat{\sigma_j^2} &= \frac{\log_g(W_j)}{N} - \widehat{\mu_j}^2= \frac{\sum_{i=1}^N \widehat{x_{i,j}}^{(2)}}{N} - \widehat{\mu_j}^2
\end{align*}
Finally, using the derived $(\widehat{\mu_j}, \widehat{\sigma_j^2})$, $\mathbb{A}$ computes the Normal $pdf$ approximation $\mathcal{N}_j \sim \mathcal{N}(\widehat{\mu_j},\widehat{\sigma_j^2})$ for each attribute $j$. 

\descr{Ranking.} In order to estimate the amount of ``valuable'' information (i.e., sensitivity) that each attribute leaks, we propose to measure the \emph{distance} (i.e., divergence) between the Normal approximation $\mathcal{N}_j$ and the Uniform distribution $\mathcal{U}$. 
This makes sense because divergence measures distance between distributions: By comparing $\mathcal{N}_j$ to the Uniform, we measure how much information $\mathcal{N}_j$ leaks compared to the distribution $\mathcal{U}$ that leaks the least amount of information~\cite{hamilton1996attribute}.  
This approach applies to a variety of computing scenarios. 
For example, a related concept was studied in~\cite{feldman1995knowledge,hilderman1999ranking} for measuring the ``interestingness'' of textual data by comparing it to an expected model, usually with the Kullback-Liebler (KL) divergence.

To the best of our knowledge, we are the first to explore this approach in the context of information privacy. Instead of the KL divergence, we rely on the Jensen-Shannon ($JS$) divergence for two reasons: (1) $JS$ is a symmetric and (2) bounded equivalent of the KL divergence. It is defined as:
\begin{align*}
JS(u,q) &= \frac{1}{2} KL(u,m) + \frac{1}{2} KL(q,m) = H(\frac{1}{2} u + \frac{1}{2} q) - \frac{1}{2} H(u) - \frac{1}{2} H(q)
\end{align*}
where $m = u/2 + q/2$ and $H$ is the Shannon entropy. As JS is in $[0,1]$ (when using the logarithm base $2$), it quantifies the relative distance between $\mathcal{N}_j$ and $\mathcal{U}_j$, and also provides absolute comparisons with distributions different from the uniform.

As JS operates on discrete values, $\mathbb{A}$ must first discretize distributions $\mathcal{N}_j$ and $\mathcal{U}_j$. Given the knowledge of intervals $\{m_j,\ldots,M_j\}$ for each attribute $j$, we can use Riemann's centered sum to approximate a definite integral, where the number of approximation bins is related to the accuracy of the approximation. We choose the number of bins to be $M_j-m_j$, and thus guarantee a bin width of $1$. We approximate $\mathcal{N}_j$ by the discrete random variable d$\mathcal{N}_j$ with the following probability mass function:
\[
\vec{Pr}(\text{d}\mathcal{N}_j) = \Spvek{Pr(x_j = m_j);Pr(x_j = m_j+1);\vdots;Pr(x_j = M_j)} = \Spvek{pdf_j(\frac{1}{2} (m_j + m_j-1));pdf_j(\frac{1}{2} (m_j+1 + m_j));\vdots;pdf_j(\frac{1}{2} (M_j + M_j-1))}
\]
where $pdf_j$ is the probability density function $\mathcal{N}_j$ and $x_j \in \{m_j,\ldots,M_j\}$. We then normalize $\vec{Pr}(\text{d}\mathcal{N}_j)$ such that $\sum_{k} Pr(x_j = k) = 1$, for each $j$. For the uniform distribution $\mathcal{U}_j$, the discretization to d$\mathcal{U}_j$ is straightforward, i.e., $\vec{Pr}(\text{d}\mathcal{U}_j) = (1/(M_j-m_j),\ldots,1/(M_j-m_j))^T$, where $\dim(\text{d}\mathcal{U}_j) = M_j-m_j$. %

$\mathbb{A}$ can now compute distances $d_j = JS(\text{d}\mathcal{N}_j,\text{d}\mathcal{U}_j) \in [0,1]$ and rank attributes in increasing order of information leakage such that $d_{\rho_1} \leq d_{\rho_2} \leq\ldots\leq d_{\rho_K}$, where $\rho_1 = \argmin_j d_j$ and $\rho_z$ (for $2 \leq z\leq K$) are defined as $\rho_z = \argmin_{\substack{j\neq \{\rho_k\}_{k = 1}^{z-1}}} (d_j)$

At this point, $\mathbb{A}$ computed the 3-tuple $(d_{\rho_j},\widehat{\mu_j},\widehat{\sigma^2_j})$ for each attribute $j$. Each user $i$ can now decide whether it is comfortable sharing attribute $j$ given distance $d_j$ and privacy sensitivity $\lambda_{i,j}$. 
To do so, each user $i$ sends $\lambda_{i,j}$ to $\mathbb{A}$ for comparison. $\mathbb{A}$ then checks which users are willing to share each attribute $j$ and updates the ratio $\gamma_j = S_j/N$, where $S_j$ is the number of users that are comfortable sharing, i.e., $S_j = |\{i \in \mathbb{U} \mbox{ s.t. } d_j \leq 1- \lambda_{i,j}\}|$. In practice, $\mathbb{A}$ could then use the majority rule to decide whether or not to monetize attribute $j$. 

\descr{Pricing.} After this ranking phase, the data broker $\mathbb{A}$ concludes the process with the pricing and revenue phases.
Prior work shows that users assign unique monetary value to different types of attributes depending on several factors, such as offline/online activities~\cite{carrascal2013your}, type of third-parties involved~\cite{carrascal2013your}, privacy sensitivity~\cite{aperjis2012market}, amount of details and fairness~\cite{malheiros2013fairly}.

We measure the value of aggregates depending on their sensitivity, the number of contributing users, and the cost of each attribute. Without loss of generality, we estimate the value of an aggregate $j$ using the following linear model:
\begin{align*}
Cost(j) = Price(j) \cdot d_j \cdot N
\end{align*}
where $Price(j)$ is the monetary value that users assign to attribute $j$. Without loss of generality, we assume in our pricing scheme a relative value of $1$ for each attribute. Existing work discussed the value of user attributes, and estimated a large range from \$ 0.0005 to \$33~\cite{carrascal2013your,olejnik2013selling} highlighting the difficulty in determining a fixed price. In practice, this is likely to change depending on the monetization scenario.

$\mathbb{A}$ then sends the set of 2-tuples $\{(d_{\rho_z},Cost(\rho_z))\}_{\rho_z = 1}^K$ to $\mathbb{C}$. Based on the tuples, $\mathbb{C}$ selects the set $\mathbb{P}$ of  attributes it wishes to purchase. After the purchase is complete, $\mathbb{A}$ re-distributes revenue $R$ among users and itself, according to the agreement stipulated with the users upon their first registration with $\mathbb{A}$. 

We consider a standard revenue sharing monetization scheme, where the revenue is split among users and the data aggregator (i.e., aggregator takes commissions):
\begin{align*}
R(\mathbb{A}) &= \sum_{\substack{j \in \mathbb{P}}} \omega_j \cdot Cost(j), \qquad R(i) = \frac{1}{N}\sum_{\substack{j \in \mathbb{P}}} (1-\omega_j) \cdot Cost(j), \qquad \forall i\in\mathbb{U}
\end{align*}
where $\omega_j$ is the commission percentage of $\mathbb{A}$. This system is popular in existing aggregating schemes~\cite{DataCommons}, credit-card payments, and online stores (e.g., iOS App Store). 
We assume a fixed $\omega_j$ for each attribute $j$.

\section{Evaluation}\label{sec:eval}
To test the relevance and the practicality of our privacy-preserving monetization solution, we measure the quality of aggregates, the overhead, and generated revenue. In particular, we study how the number of protocol participants and their privacy sensitivities affect the accuracy of the Gaussian approximations, the computational performance, the amount of information leaked for each attribute, and revenue. 

\subsection{Setup}\label{subsec:dataset}
We consider secret shares in $\mathbb{Z}_p$ where $p$ is a  1024 bits modulus, the number of users $N \in [10,100000]$, and each user $i$ with profile $\vec{p_i}$. %
We implemented our privacy-preserving protocol in Java, and rely on public libraries for secret key initialization, for multi-threading decryption, and on the MALLET~\cite{McCallumMALLET} package for computation of the JS divergence.

We run our experiments on a machine equipped with Mac OSX 10.8.3, dual-core Core i5 processor, 2.53 GHz, and 8 GB RAM. Measurements up to 100 users are averaged over 300 iterations, and the rest (from 1k to 100k users) are averaged over 3 iterations due to large simulation times.

We populate user profiles with U.S. Census Bureau information~\cite{dataferrett,datagov}: We obtained anonymized offline and online attributes about 100,000 people. We pre-processed the acquired data by removing incomplete profiles (i.e., some respondents prefer not to reveal specific attributes). 

Without loss of generality, we focus on three types of offline attributes: Yearly \emph{income} level, \emph{education} level and \emph{age}. We selected these attributes because (1) a recent study~\cite{carrascal2013your} shows that these attributes have high monetary value (and thus privacy sensitivity), and (2) they have significantly different distributions across users. This allows us to compare retribution models, and measure the accuracy of the Gaussian approximation for a variety of distributions. 

Table \ref{table:datasets} shows the mean and standard deviation for the three considered attributes with a varying number of users. Note that the provided values for \emph{income} and \emph{education} use a specific scale defined by the Census Bureau. For example, a value of $1$ and $16$ for \emph{education} correspond to ``Less than 1st grade'' and ``Doctorate'', respectively. 

We could consider other types of attributes as well, such as internet, music and video preferences from alternative sources, such as Yahoo Webscope~\cite{webscope}. Although an exhaustive comparison of the monetization of all different attributes is an exciting perspective, it is out of the scope of this paper and we leave this for future work. %

\begin{table}[t]
\centering
\includegraphics[width=0.65\linewidth]{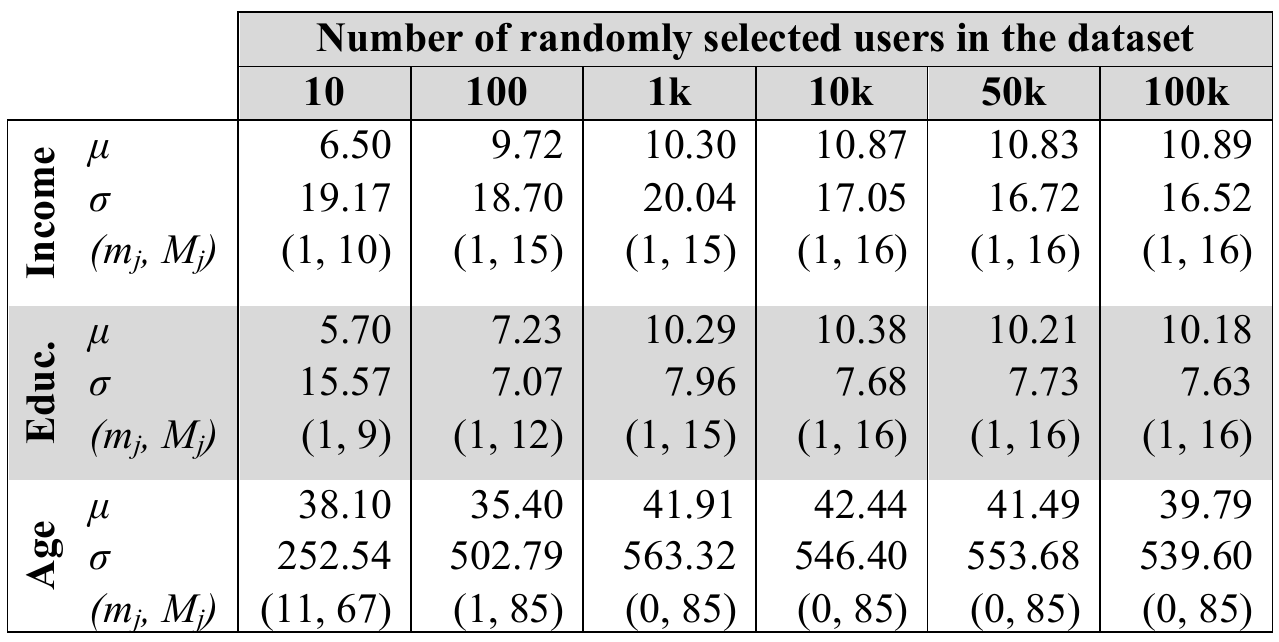}
\caption{Summary of the U.S. Census dataset used for the evaluation. We considered three types of attributes (level of income, education and age), which reflect different types of sample distributions (as shown in Fig. \ref{fig:gaussApprox}).}
\label{table:datasets}
\vspace{-0.15cm}
\end{table}

\begin{figure*}[t]
\centering
\begin{subfigure}[b]{\textwidth}
                \includegraphics[width=\textwidth]{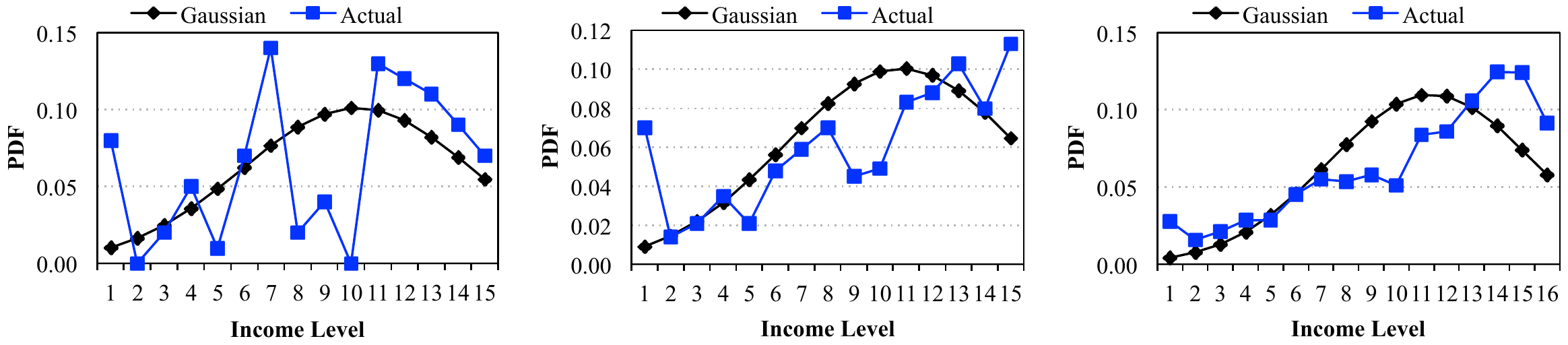}
                \caption{Attribute \emph{income}, sampled from 100 users (left), 1k users (middle) and 100k users (right).}
                \label{fig:gaussApproxIncome}
                        \end{subfigure}\\\vspace{0.2cm}
        \begin{subfigure}[b]{\textwidth}
                \includegraphics[width=\textwidth]{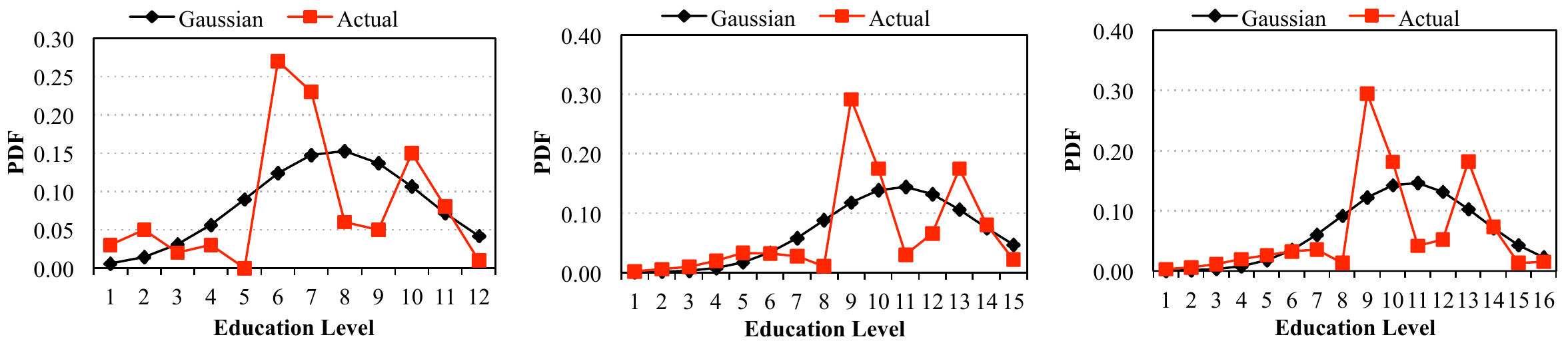}
                \caption{Attribute \emph{education}, sampled from 100 users (left), 1k users (middle) and 100k users (right).}
                \label{fig:gaussApproxEduca}
        \end{subfigure}\\\vspace{0.2cm}
        \begin{subfigure}[b]{\textwidth}
                \includegraphics[width=\textwidth]{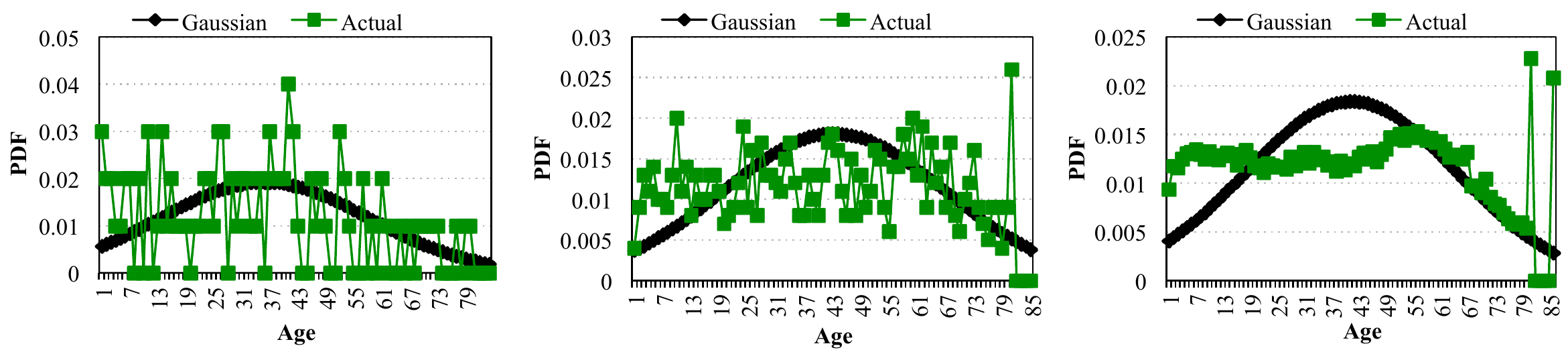}
                \caption{Attribute \emph{age}, sampled from 100 users (left), 1k users (middle) and 100k users (right).}
                \label{fig:gaussApproxAge}
        \end{subfigure}%
                \vspace{-0.1cm}
                \caption{Gaussian approximation vs. actual distribution for each considered attribute.}
                \label{fig:gaussApprox}
\vspace{-0.1cm}
\end{figure*}

\subsection{Results}

We evaluate four aspects of our privacy-preserving scheme: model accuracy, information leakage, overhead and pricing. 

\subsubsection{Model Accuracy}

In our proposal, we approximate empirical probability density functions with Gaussian distributions. The accuracy of approximations is important to assess the relevance of derived data models. In Fig. \ref{fig:gaussApprox}, we compare the actual distribution of each attribute with their respective Gaussian approximation and vary the number of users from 100 to 100,000. Note that in order to compare probabilities over the domain $[m_j, M_j]$, we scaled both the actual distribution and the Gaussian approximation such that their respective sums over that domain are equal to one. We observe that, visually, the Gaussian approximation captures general trends in the actual data. 

\begin{figure*}[t]
\centering
\begin{subfigure}[b]{0.43\textwidth}
                \includegraphics[width=\linewidth]{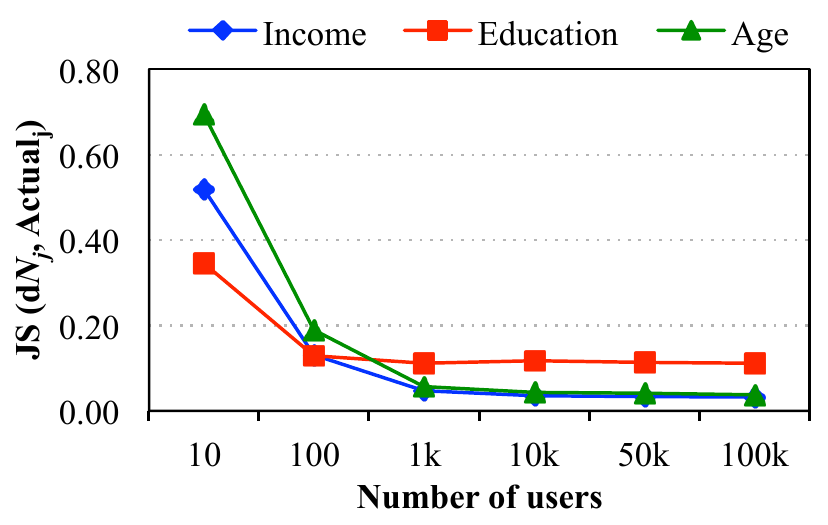}
                \caption{Divergence between the Gaussian approximation and the actual distribution of each attribute $j$, computed as the $JS(\text{d}\mathcal{N}_j,\text{Actual}_j)$. Lower values indicate better accuracy.}
                \label{fig:distToActual}
                        \end{subfigure}\hspace{15mm}
\begin{subfigure}[b]{0.43\textwidth}
                \includegraphics[width=\linewidth]{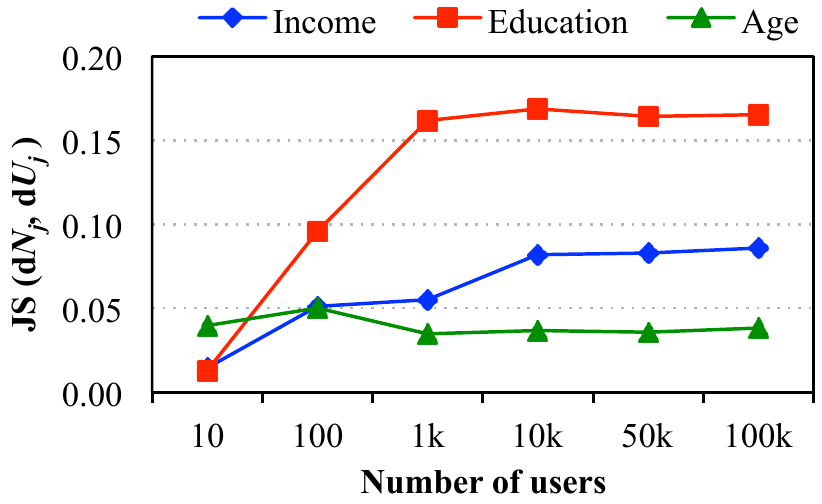}
                \caption{Information leakage for each type of attribute $j$ (income, education and age), defined as $JS(\text{d}\mathcal{N}_j,\text{d}\mathcal{U}_j)$. Lower values indicate smaller information leaks.}
                \label{fig:infoLeakage}
                        \end{subfigure}
\begin{subfigure}[b]{0.43\linewidth}
                \includegraphics[width=\linewidth]{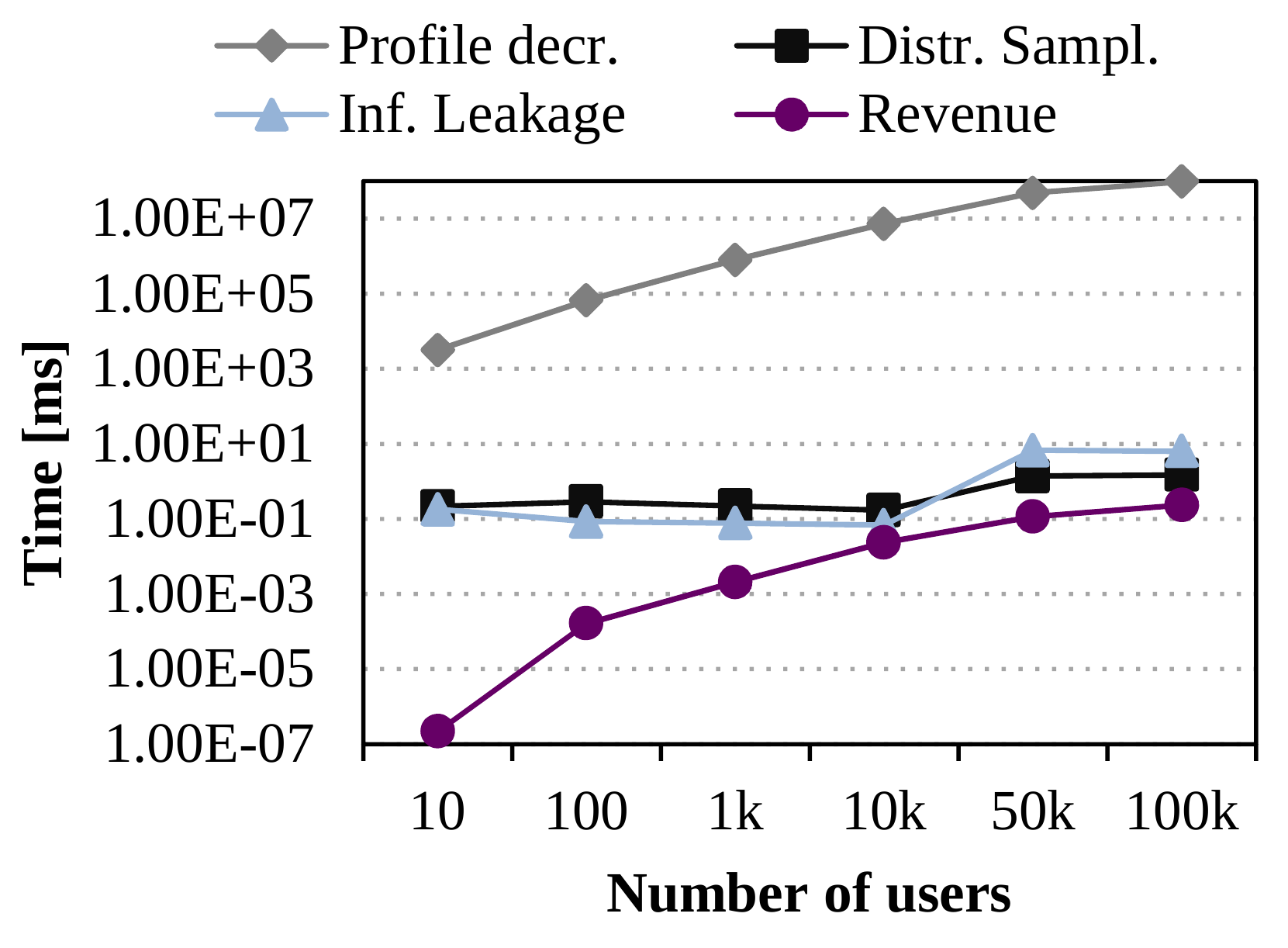}
                \caption{Performance measurements for each of the four phases of the protocol performed by the data broker.}
                \label{fig:perfBroker}
                        \end{subfigure}\hspace{15mm}
\begin{subfigure}[b]{0.43\textwidth}
                \includegraphics[width=\linewidth]{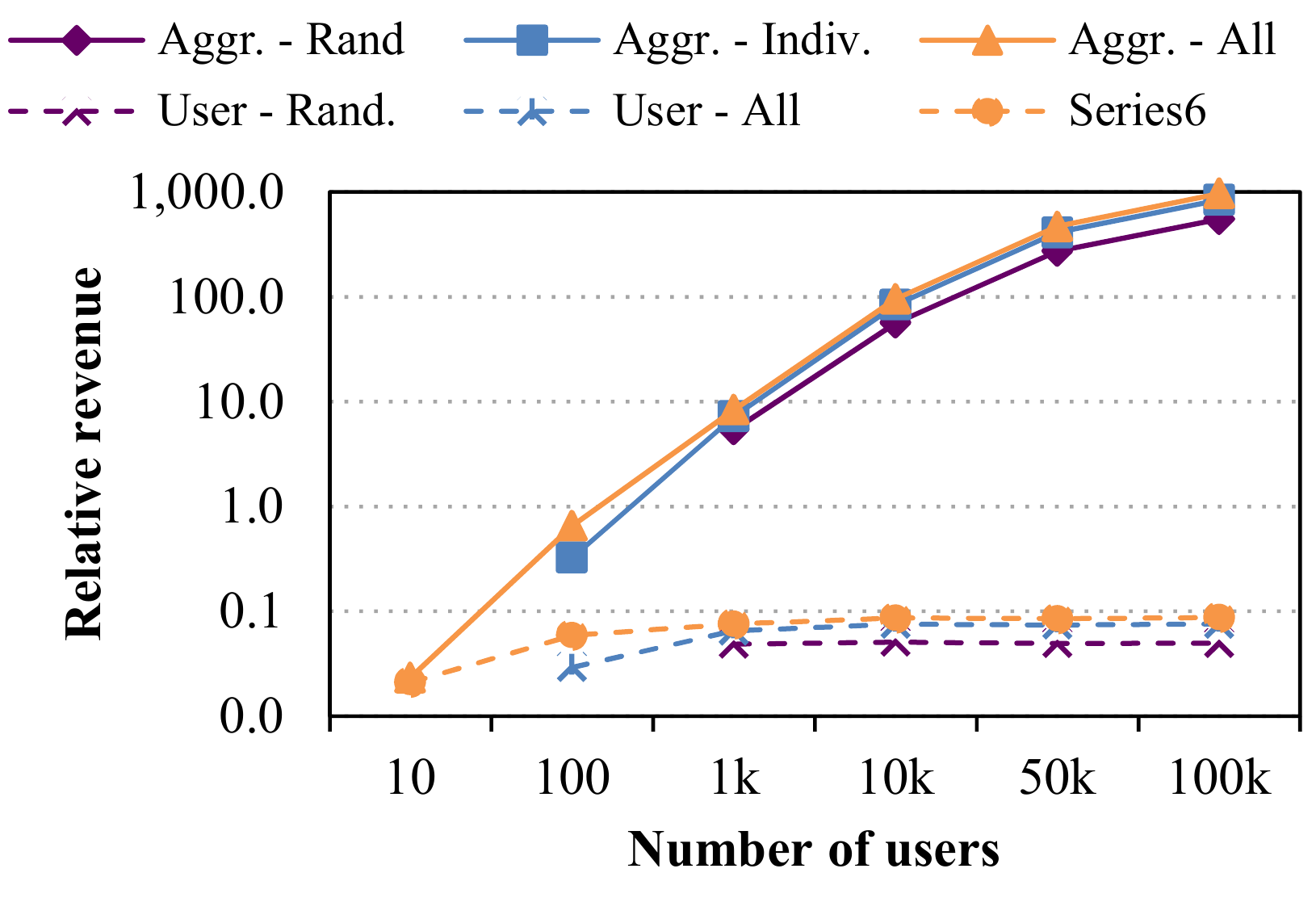}
                \caption{Relative revenue (per attribute) for each user $i \in \mathbb{U}$ and the data aggregator $\mathbb{A}$, assuming that an attribute is valued at $1$.}
                \label{fig:remuneration}
                        \end{subfigure}
                        \label{fig:results}
                        \caption{Results of the evaluation of the proposed framework on the U.S. Census dataset.}
\end{figure*} 

We measure the accuracy of the Gaussian approximation in more details with the JS divergence (Fig. \ref{fig:distToActual}). We observe that with 100 users, the approximation reaches a plateau for \emph{education}, whereas \emph{income} and \emph{age} require 1k users to converge. For the two latter attributes, the approximation accuracy triples when increasing from 100 to 1k users. Moreover, as the number of user increases, the fit of the Gaussian model for \emph{income} and \emph{age} is two times better (JS of 0.05 bits) than for \emph{education} (JS of 0.1 bits). The main reason is that \emph{education} has more data points with large differences between actual and approximated distributions than \emph{income} and \emph{age} (as shown in Fig. \ref{fig:gaussApprox}). 

These results indicate that, for non-uniform distributions, the Gaussian approximation is accurate with a relatively small number of users (about 100). 
It is interesting to study this result in light of the Central Limit Theorem (CLT). Remember that the CLT states that the arithmetic mean of a sufficiently large number of variables will tend to be normally distributed. In other words, a Gaussian approximation quickly converges to the original distribution and this confirms the validity of our experiments.  
This also means that $\mathbb{C}$ can obtain accurate models even if it requests aggregates about small groups of users. 
In other words, collecting data about more than 1k users does not significantly improve the accuracy of approximations, even for more extreme distributions. %

\subsubsection{Information Leakage}\label{subsec:leakage}
We compare the divergence between Gaussian approximations and uniform distributions to measure the information leakage of different attributes. Fig. \ref{fig:infoLeakage} shows the sensitivity for each attribute with a varying number of users. We observe that the amount of information leakage stabilizes for all attributes after a given number of participants. In particular, \emph{education} and \emph{age} reach a maximum information leakage with 1k users, whereas 10k users are required for \emph{income} to achieve the same leakage. 

Overall, we observe that \emph{education} is by far the attribute with the largest distance to the uniform distribution, and therefore arguably the most valuable one. In comparison, \emph{Income} and \emph{age} are 50\% and 75\% less ``revealing''. Information leakage for \emph{age} decreases from 100 to 1k users, as age distribution in our dataset tends towards a uniform distribution. In contrast, \emph{education} and \emph{income} are significantly different from a uniform distribution. An important observation is that the amount of valuable information does not increase monotonically with the number of users: For \emph{age}, it decreases by 30\% when the number of users increases from 100 to 1k, and for \emph{education} it decreases by 3\% when transitioning from 10k to 50k users. 

These findings show that larger user samples do not necessarily provide better discriminating features. 
This also shows that users should not decide whether to participate in our protocol solely based on a fixed threshold over total participants, as this may prove to leak slightly more private information. 

\subsubsection{Overhead}\label{subsec:performance}
We measure the computation overhead for both users and the data broker. For each user, we find that one execution of the protocol requires 0.284 ms (excluding communication delays), out of which 0.01 ms are spent for the profile generation, 0.024 ms for the feature extraction, 0.026 ms for the differential-privacy noise addition, and 0.224 ms for encryption of the noisy attribute. In general, user profiles are not subject to change within short time intervals, thus suggesting that user-side operations could be executed on resource-constrained devices such as mobile phones. %

From Fig. \ref{fig:perfBroker}, observe that the data broker requires about one second to complete its phases when there are only 10 users, 1.5 min with 100 users, 15 min with 1k users, and 27.7 h for 100k users. Note, however, that running times can be remarkably reduced using algorithmic optimization and parallelization, which is part of our future work. In our results, decryption is the most time-consuming operation for the data broker as it incurs ($O(N\cdot M_j)$): this could be reduced to $O(\sqrt{N\cdot M_j})$ by using the Pollard's Rho method for computing the discrete logarithm \cite{pollard1978monte}. Also, decryption can be speedup up by splitting decryption operations across multiple machines (i.e., the underlying algorithm is highly-parallelizable). 

\subsubsection{Pricing}\label{subsec:pricing-remun}
Recall that the price of an attribute aggregate depends on the number of contributing users, the amount of information leakage, and the cost of the attribute. 
We consider that each attribute $j$ has a unit cost of $1$ and the data broker takes a commission $\omega_j$. %
We consider three types of privacy sensitivities $\lambda$:
(i) a uniform random distribution of privacy sensitivities $\lambda_{i,j}$ for each user $i$ and for each attribute $j$, 
(ii) an individual privacy sensitivity $\lambda_i$ for each user (same across different attributes), and 
(iii) an all-share scenario ($\lambda_i = 0$ and all users contribute). 
The commission percentage is set to $\omega_j = \omega = 0.1$.

Fig. \ref{fig:remuneration} shows the average revenue generated from one attribute by the data broker and by users. 
We observe that user revenue is small and does not increase with the number of participants. %
In contrast, the data broker revenue increases linearly with the number of participants. %
In terms of privacy sensitivities, we observe that with higher privacy sensitivities ($\lambda_i > 0$), fewer users contribute, thus generating lower revenue overall and per user. For example, users start earning revenue with 10 participants in the all-share scenario, %
but more users are required to start generating revenue if users adopt higher privacy sensitivities. %

We observe that users are incentivized to participate as they earn some revenue (rather than not benefiting at all), but the generated revenue does not generate significant income, thus, it might encourage user participation from ``biased'' demographics (e.g., similar to Amazon Mechanical Turk). 
In contrast, the data broker has incentives to attract more users, as it revenue increases with the number of participants. 
However, customers are incentivized to select fewer users because cost increases with the number of users, and 100 users provide as good an aggregate as 1000 users. This is an intriguing result, as it encourages customers to focus on small groups of users representative of a certain population category.

\subsection{Security}\label{subsec:privacy}
\descr{Passive adversaries.}
To ensure privacy of the personal user attributes, our framework relies on the  security of the underlying encryption and differential-privacy methods presented in \cite{shi2011privacy}. Hence, no passive adversary (a user participating in the monetization protocol, the data aggregator or an external party not involved in the protocol) can learn any of the user attributes, assuming that the key setup phase has been performed correctly and that a suitable algebraic group (satisfying the DDH assumption) with a large enough prime order (1024 bits or more) has been chosen.

\descr{Active adversaries.}
As per~\cite{shi2011privacy}, our framework is resistant to collusion attacks among users and between a subset of users and the data broker, as each user $i$ encrypts its attribute values with a unique and secret key $s_i$. However, pollution attacks, which try to manipulate the aggregated result by encrypting out-of-scope values, can affect the aggregate result of our protocol. Nevertheless, such attacks can be mitigated by including, in addition to encryption, range checks based on efficient (non-interactive) zero-knowledge proofs of knowledge \cite{blum1988non,boudot2000efficient,lipmaa2003diophantine}: each user could submit, in addition to the encrypted values, a proof that such values are indeed in the plausible range specified by the data aggregator. However, even within a specific range, a user can manipulate its contributed value and thus affect the aggregate. Although nudging users to reveal their true attribute value is an important challenge, it is outside of the scope of this paper. %

\section{Related Work}\label{sec:rel-work}
Our work builds upon two main domains, in order to provide the privacy and incentives for the users and data aggregators: (1) privacy-preserving aggregation~\cite{erkin2012private,shi2011privacy,shi2010prisense,xing2013mutual}, and (2) privacy-preserving monetization of user profiles~\cite{backes2012obliviad,guha2011privad,riederer2011sale,toubiana2010adnostic}. Hereafter we discuss these two sets of works.

\subsection{Privacy-Preserving Aggregation}\label{data-aggr}
Erkin and Tsudik~\cite{erkin2012private} design a method to perform privacy-preserving data aggregation in the smart grid. Smart meters jointly establish secret keys without having to rely on a trusted third party, and mask individual readings using a modified version of the Paillier encryption scheme~\cite{paillier1999public}. The aggregator then computes the sum of all readings without seeing individual values. %
Smart meters must communicate with each other, thus limiting this proposal to online settings. Shi et al.~\cite{shi2010prisense} compute the sum of different inputs based on data slicing and mixing with other users, but have the same limitation: all participants must actively communicate with each other during the aggregation. 

Another line of work~\cite{chen2013splitx,chen2012towards} introduces privacy-preserving aggregation by combining homomorphic encryption and differential privacy, i.e., users encrypt their data with the customer public key and send it to a trusted aggregator. The aggregator adds differential noise to encrypted values (using the homomorphic property), and forwards the result to the customer. The customer decrypts contributions and computes desired aggregates. 
These proposals, however, suffer from a number of shortcomings as: (i) they rely on a trusted third party for differential privacy;  (ii) they require at least one public key operation per single bit of user input, and one kilobit of data per single bit of user answer, or rely on XOR encryption; and (iii) contributions are linkable to users as aggregation occurs after decryption.

The work by Shi et al.~\cite{shi2011privacy} supports computing the sum of different inputs in a privacy-preserving fashion, without requiring communication among users, nor repeated interactions with a third party. It also provides differential privacy guarantees in presence of malicious users, and establishes an upper bound on the error induced by the additive noise. This work formally shows that a Geometric distribution provides ($\epsilon,\delta$)-differential privacy (DD) in $\mathbb{Z}_p$. %
We extend the construction in~\cite{shi2011privacy} to support the privacy-preserving computation of probability distributions (in addition to sums). Intuitively, we use the proposed technique to compute the parameters of Gaussian approximations in a privacy-preserving way. As we maintain the same security assumptions, our framework preserves provable privacy properties. As part of future work, we intend to explore the properties of regression modeling and privacy-preserving computation of regression parameters \cite{ahmadi2010privacy,xing2013mutual}, in addition to distributions.

\subsection{Privacy-Preserving Monetization}\label{subsec:beh-adv}
Previous work investigated two main approaches to privacy-preserving  Online Behavioral Advertisement (OBA). 
The first approach minimizes the data shared with third parties, by introducing local user profile generation, categorization, and ad selection~\cite{akkus2012non,guha2011privad,mohan2013prefetching,toubiana2010adnostic}. The second approach relies on anonymizing proxies to shield users' behavioral data from third parties, until users agree to sell their data \cite{backes2012obliviad,riederer2011sale}. 

Toubiana et al.~\cite{toubiana2010adnostic} propose to let users maintain browsing profiles on their device and match ads with user profiles, based on a cosine-similarity measure between visited websites meta-data (title, URL, tags) and ad categories. Users receive a large number of ads, select appropriate ones, and share selected ads with ad providers (not revealing visited websites nor user details). Guha et al.~\cite{guha2011privad} propose to do the ad matching with an anonymization proxy instead. %
Although the cost of such system is estimated at \$0.01/user per year, such solution demands significant changes from web browser vendors and online advertisers.
Akkus et al. propose to let users rely on the website publisher to anonymize their browsing patterns \emph{vis-\`a-vis} the ad-provider. Their protocol introduces significant overhead: The website publisher must repeatedly interact with each visitor and forward encrypted messages to the ad-provider.

Instead of local profiles, Riederer et al.~\cite{riederer2011sale} propose a fully centralized approach, where an anonymization proxy mediates interactions between users and website publishers. The proxy releases the mapping between IP addresses and long-term user identifiers only after users agree to sell their data to a customer, thus allowing the customer to link different visits by the same users. %
However, users have to entrust a third party with their personal information.

In contrast, our framework does not rely on any additional user-side software, does not impose computationally expensive cryptographic computation on user devices, and prevents the customer from learning individual user data.

\section{Conclusion}\label{sec:concl}
As the amount and sensitivity of personal data shared with service providers increases, so do privacy concerns. 
Users usually have little control over what information is processed by service providers and how it is monetized with advertisers. Our work offers a privacy-preserving alternative where users only disclose an aggregate model of their profiles, by means of encrypted and differentially private contributions.
Our solution tackles trust and incentive challenges: rather than selling data \emph{as-is}, users trade a \emph{model} of their data. 
Users also monetize their profiles by dynamically assessing the value of data aggregates. To this end, we use an information-theoretic measure to compute the amount of valuable information provided to advertisers.

We evaluate our framework on a real and anonymized dataset with more than 100,000 users (obtained from the U.S. Census Bureau) and show, with an experimental evaluation, that our solution (i) provides accurate aggregates with as little as 100 users, (ii) introduces low overhead for both users (less than 1ms on commodity hardware) and data aggregators, and (iii) generates revenue for both users and aggregators. 

As part of future work, we plan to enhance our scheme with new features, including fault-tolerant aggregation~\cite{chan2012privacy}, which can be integrated in order to allow users to join/leave dynamically without disrupting the scheme. Also, range checks for the encrypted user attributes, based on efficient zero-knowledge proofs, could thwart active pollution attacks. 
Users could also contribute higher order moments (e.g., $x^3$ or $x^4$) for the aggregator to obtain more precise approximations using moment-generating functions (an alternative to $pdf$s). 
Finally, we intend to investigate schemes for targeting ads to users contributing data to the aggregation, by allowing the aggregator to select specific subgroups of users according to the customer's target population.

\bibliographystyle{abbrv}
\bibliography{biblio}  %

\end{document}